\documentclass{emulateapj}

\def\ga{\mathrel{\hbox{\rlap{\hbox{\lower4pt\hbox{$\sim$}}}\hbox{$>$}}}}
\def\la{\mathrel{\hbox{\rlap{\hbox{\lower4pt\hbox{$\sim$}}}\hbox{$<$}}}}

\usepackage{verbatim}
\usepackage{amsmath}
\usepackage{multirow}
\usepackage{relsize}
\usepackage{amssymb}
\usepackage{float}

\newcommand{\alphavir}{\alpha_\mathrm{vir}}

\newcommand{\sfrff}{\mathrm{SFR}_\mathrm{ff}}

\newcommand{\sigs}{\sigma_s}
\newcommand{\scrit}{s_\mathrm{crit}}
\newcommand{\rhocrit}{\rho_\mathrm{crit}}
\newcommand{\eps}{\epsilon_0}
\newcommand{\phit}{\phi_t}

\newcommand{\tff}{t_\mathrm{ff}}

\newcommand{\meanrho}{\rho_0}

\newcommand{\means}{s_0}
\newcommand{\deriv}{\,\mathrm{d}}

\newcommand{\ycut}{y_\mathrm{cut}}

\newcommand{\s}{\mathrm{s}}

\shorttitle{The Accelerating Star Formation Rate}

\shortauthors{Burkhart}
\begin{document}
\title{The Star Formation Rate in the Gravoturbulent Interstellar Medium}
\author{Blakesley Burkhart\altaffilmark{1}}
\altaffiltext{1}{Harvard-Smithsonian Center for Astrophysics, 60 Garden st. Cambridge, MA, 02138,  USA}

\begin{abstract}
Stars form in supersonic turbulent molecular clouds that are self-gravitating. 
We present an analytic determination of the star formation rate (SFR) in a gravoturbulent medium based on the density probability distribution function of molecular clouds having a piecewise lognormal \textit{and} power law  form.
This is in contrast to previous analytic SFR models that are governed primarily by interstellar turbulence which sets purely lognormal density PDFs.
In the gravoturbulent SFR model described herein, low density gas resides in the lognormal portion of the PDF. Gas becomes gravitationally unstable past a critical density ($\rho_{crit}$), and the PDF begins to forms a power law. As the collapse of the cloud proceeds, the transitional density ($\rho_t$) between the lognormal and power law portions of the PDF moves towards lower-density while the slope of the power law ($\alpha$) becomes increasingly shallow.
The star formation rate per free-fall time is calculated via an integral over the lognormal from  $\rho_{crit}$ to $\rho_t$ and an integral over the power law from $\rho_t$ to the maximum density. As $\alpha$ becomes shallower the SFR accelerates beyond the expected values calculated from a lognormal density PDF.
We show that the star formation efficiency per free fall time in observations of local molecular cloud increases with shallower PDF power law slopes, in agreement with our model. 
Our model can explain why star formation is spatially and temporally variable within a cloud and why the depletion times observed in local and extragalactic giant molecular clouds vary.  Both star-bursting and quiescent star-forming systems can be explained without the need to invoke extreme variations of turbulence in the local interstellar environment.

\end{abstract}

\section{Introduction}
\label{intro}
Stars are the fundamental link between galaxy evolution, planet formation, and astrobiology \citep{chyba2005,Mckee_Ostriker2007,krumreview2014}.  Stars inject energy and momentum into the surrounding interstellar and intergalactic medium, an effect commonly known as stellar feedback, via winds and supernova \citep{Fall10a,Agertz15a,Lopez2011ApJ...731...91L,Rosen2014,Rosen2016,hayward17}. 
Despite the importance of stars for galaxy evolution, a full observational and theoretical understanding of their formation remains elusive. If gravity alone was relevant, galactic interstellar clouds with density $\rho$ should collapse to form stars on roughly a free-fall time:
 \begin{equation}
 t_{ff}=\sqrt{3\pi/(32 G \rho)}~~;
 \label{ff}
 \end{equation}
 
However the dynamics of the star forming molecular interstellar medium are complicated by energetics that can counteract gravity, including magnetic fields and supersonic turbulence \citep{Mouschovias76a,Lars81,Shu87a,McK93,MacLow2004,Mckee_Ostriker2007,Lazarian07rev,Kowal2009,Crutcher09a,Crutcher10a,Crutcher12a,LEC12,Mocz2017},
and the timescales for star formation are consequently much longer than the free-fall time.  Additional factors that complicate our understanding of star formation include the many orders of magnitude in spatial and temporal scales and observational biases such as line-of-sight effects, limited resolution and incomplete density tracers \citep{goodman09,Beaumont2015,Lombardi2015AA,schneider2015MNRAS.453L..41S,Chen2017,Alves2017AA}.

One of the main challenges for a theory of star formation is 
determining the star formation rate (SFR) and star formation efficiency (SFE) in both local molecular clouds and in galaxies \citep{krumreview2014}.  The current paradigm for the initial conditions of star formation is that stars form in dense filamentary molecular clouds that result from gravo-magnetohydrodyanmic (MHD)-turbulent fragmentation \citep{maclow04,Krumholz2005,Padoan11b,Hennebelle11a,federrath12,Federrath13b,Collins12a,Hopkins13b,Burkhart12,Lee2015ApJ...800...49L,Semenov2016,Mocz2017,Semenov2017}.  The initial conditions imprinted on the diffuse
and molecular gas on parsec scales (e.g., the level of
turbulence, the cloud density, and the structure of the magnetic
field) may determine the key properties of the initial
mass function (IMF) and the star formation rates in
galaxies \citep{krumreview2014}.

In the absence of self-gravity, MHD turbulence sets a lognormal (LN) initial density distribution in star-forming clouds \citep{Vazquez-Semadeni1994,Padoan1997,Scalo98a}:
\begin{equation}
p_{LN}(s)=\frac{1}{\sqrt{2\pi\sigs^2}}\exp\left(-\frac{(s-s_0)^2}{2\sigs^2}\right)\,,
\label{eq:LNpdf}
\end{equation}
expressed in terms of the logarithmic density,
\begin{equation} \label{eq:s}
s\equiv\ln{(\rho/\meanrho)}\,.
\end{equation} and where $\sigs$ is the standard deviation of the lognormal.
The quantities $\meanrho$ and $\means$ denote the mean density and mean logarithmic density\footnote{We note that $s_0$ should not be confused as the logarithmic density
evaluated at $\rho_0$.}, the latter of which is related to $\sigs$ by
\begin{equation}
\means=-\frac{1}{2}\,\sigs^2
\end{equation}

Analytic calculations for the star formation rate have used the lognormal PDF \citep{Krumholz2005,Padoan11b,Hennebelle11b,federrath12,Renaud12a,Hopkins12d,Gribel2017}, taking the integral starting from a critical density for collapse ($\rho_{crit}$), which varies for different authors. In these works, the SFR depends on the exact choice of a number of parameters of order unity as well as on the properties of the lognormal PDF, set by supersonic turbulence.  In particular the width of the lognormal PDF is controlled by the sonic Mach number ($M_s$) and a dimensionless turbulent forcing parameter ($b$, \citet{Federrath2008}):
\begin{equation}
\sigs^2=\ln[1+b^2M_s^2]
\label{eqn.sigma}
\end{equation}
Deviations from the lognormal form have been observed but do not strongly affect the SFR calculation. For example, if the gas equation of state is non-isothermal, the SFR calculation can decrease by roughly a factor of a few \citep{federrath2015MNRAS.448.3297F}.

Simulations of \textit{self-gravitating MHD turbulence} suggest that the gas density PDF stems from a combination of turbulence, which dominates the dynamics at low densities, and self-gravity, which dominates at high densities. 
Gravitationally unstable free-falling density structures are characterized
by a power-law (PL) PDF \citep{Ballesteros-Paredes11a,Collins12a,Girichidis2014,MyersP2015,Burkhart2017ApJ...834L...1B,Mocz2017,padoan2017ApJ...840...48P,MyersP2017}: 

\begin{align}
p_{PL}(s) = 
C e^{-\alpha s}, \;\;\; & s > s_t ,
\label{eqn.PL}
\end{align}

\noindent 
where $s_t=\rm{ln}(\rho_t/\rho_0)$ is the logarithm of the normalized transitional density between the lognormal and power law forms of the density PDF.
Observational studies of giant molecular clouds (GMCs) have confirmed that the highest column
density regime (corresponding to visual extinction A$_V >1$) of the PDF often has a power law distribution while the lower column density material in the PDF is well-described by a lognormal form \citep{Kainulainen09a,Lombardi10a,schneider2015MNRAS.453L..41S,Kainulainen13b,Hennebelle11a,federrath12,Stutz2015A&A...577L...6S,Lombardi2015AA,burkhart2015,Imara2016,Bialy2017ApJ...843...92B}.

We explore an analytic determination of the star formation rate based on the density PDF having a \textit{piecewise lognormal and power law form}. This paper is motivated by the observational and numerical evidence discussed above for the existence of a power law PDF at high densities as well as observational/numerical signatures of accelerating star formation rates \citep{Palla00a,Lee2015ApJ...800...49L,Lee2016,Caldwell2018MNRAS.474.4818C}.  In particular, \citet{Palla00a} used pre-main-sequence evolutionary tracks to show that the star formation rate has accelerated rapidly in a number of nearby GMCs over a time scale of roughly $10^7$yrs. \citet{Lee2016} showed that  variations in the star formation efficiency observed in molecular clouds can not be explained by the rather small fluctuations in the cloud turbulent velocity dispersions while \cite{Caldwell2018MNRAS.474.4818C} has shown the star formation efficiency grows quadratically in time. These observations can not be easily explained by the lognormal density star formation theories which predict constant star formation rates for given turbulence parameters.
Moreover, \citet{Leroy2017} found that in the M51 galaxy, as observed by the PAWS survey, 
the star formation efficiency per free fall time seems to be anti-correlated with the velocity dispersion, demonstrating further tension with turbulent star
formation models. 
However, as we will show, the inclusion of a time varying power law density PDF can account for both accelerated and enhanced star formation without needing to invoke extreme variations in interstellar turbulence.  

This paper is organized as follows: in Section \ref{review} we review the lognormal density PDF calculations of the SFR. In Section \ref{review2} we review the analytic derivations of \citet{Collins12a} and \citet{Burkhart2017ApJ...834L...1B} for the piecewise lognormal plus powerlaw (LN+PL) PDF transition density and normalization.  Then in Section \ref{SFR} we calculate the SFR of the LN+PL PDF and compare this calculation to the lognormal PDF models described in Section \ref{review}.  We compare the LN+PL model to observations in Section \ref{obs}. Finally we discuss our results in Section \ref{discussion} followed by our conclusions in Section \ref{conclusions}.
\begin{figure}
\includegraphics[width=9.8cm]{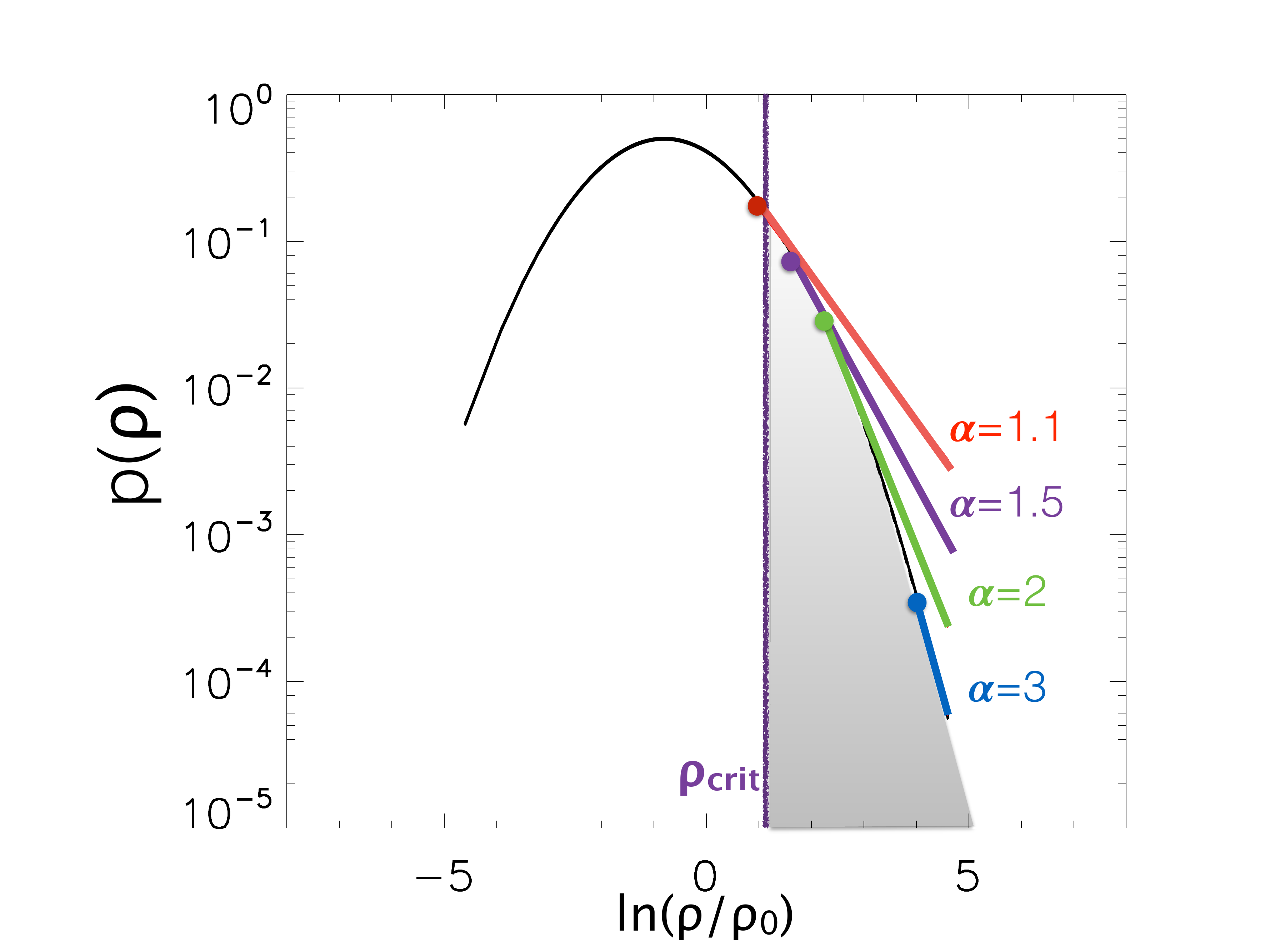}
\caption{
\label{fig:PDF_theory}
A cartoon of the density/column density PDF in and around star forming molecular clouds which consists of a lognormal distribution at low density and a power law distribution at high density. As dense material collapses, the power law slope ($\alpha$) becomes shallower and the transition point between the lognormal and power law moves towards lower density values, as indicated with different colored slopes and transition density points. The $SFR_{ff}$ is calculated as an integral over densities greater than $\rho_{crit}$, as indicated in the shaded gray box for the black lognormal PDF. }
\end{figure}

\section{Review of The Star Formation Rate based on Lognormal Density PDFs}
\label{review}

A number of past studies have developed an analytic derivation of the SFR based on the statistics of supersonic magnetized turbulence \citep{Krumholz2005,Padoan11b,Hennebelle11b,federrath12}. All of these models take an integral over the lognormal density PDF (Equation \ref{eq:LNpdf}) in order to estimate the gas mass above a given density threshold that forms stars.  
Here we briefly review the basic points of these works, namely how the SFR is derived from the lognormal PDF and its dependencies on the properties of turbulence.  We note that many other LN SFR models exist in the literature but we focus our comparison on a subset of them as the basic features are similar. The methodology outline in this section will be repeated for the LN+PL density PDF model derived in this work.

The analytic models of \citet{Krumholz2005}, \citet{Padoan11b}, and \citet{Hennebelle11b} for the SFR per free-fall ($\sfrff$) integrate the lognormal density PDF, Equation~(\ref{eq:LNpdf}) multiplied by $\rho/\meanrho$ and a normalized free fall factor. 
This then provides the \emph{star formation rate per free-fall time} ($\sfrff$), first described in \citet{Krumholz2005}, which is the mass fraction going into stars per free-fall time:
\begin{equation}
\rm{SFR_{ff}} = {\eps} \mathlarger{\int}_{\scrit}^{\infty}{\frac{\tff(\meanrho)}{\tff(\rho)} \frac{\rho}{\meanrho} \, p_{LN}(s) \deriv s}\,.
\label{eq:sfrff}
\end{equation}
where $\epsilon_0$ is the local efficiency for converting gas into stars and depends on the properties of feedback and $s_{\rm{crit}}=\rm{ln}(\rho_{crit}/\rho_0)$ is the critical density for collapse. Given the SFR per free-fall time one can also calculate the SFR:
\begin{equation}
\label{Sfrtrue}
SFR=\frac{M_{\rm{cloud}}}{t_{\rm{ff}}(\rho_0)}SFR_{ff}.
\end{equation}

There are two main distinction between different LN SFR models: the free fall factor used, i.e. the treatment of the  $\frac{\tff(\meanrho)}{\tff(\rho)}$ term in Equation \ref{eq:sfrff}, and the definition of the critical density ($s_{\rm{crit}}$).
\citet{Krumholz2005} estimate the $\sfrff$ with a factor $\tff(\meanrho)/\tff(\meanrho)=1$  while \citet{Padoan11b}  uses a factor $\tff(\meanrho)/\tff(\rhocrit)$.  Both of these studies use a free fall factor which is independent of density and can therefore be taken out of the integral.  $\tff(\meanrho)/\tff(\rho)$ appears inside the integral in studies by \citet{Hennebelle11b} and \citet{federrath12} because gas with different densities has different free-fall times. These studies are therefore termed ``multi-free fall" SFR models.
As for the differences in the critical density,  we summarize the different critical densities and corresponding SFR$_{ff}$ for the models that employ a LN density PDF (see also Table 1 of \citet{federrath12}) .

\begin{itemize}
\item \citet{Krumholz2005}:

The critical density for  Equation \ref{eq:sfrff} for the Krumholz \& Mckee (2005, KM) model is their Equation 27:
\begin{equation}
\rho_{\rm{crit,KM}}/\rho_0=\frac{\pi^2}{15}\phit^2\alphavir M_s^2
\end{equation}
where $\alphavir$ is the virial parameter, $\phi_t$ is of order unity and accounts for the uncertainty in the timescale factor originally introduced in \citet{Krumholz2005}, and $M_s$ is the sonic Mach number.
 The corresponding SFR$_{ff}$ with  $\tff(\meanrho)/\tff(\meanrho)=1$ is:

\begin{align}
\label{sfrff_km}
SFR_{\rm{ff,KM}}=\frac{\eps}{2\phit} \left\{1+\mathrm{erf}\left[\frac{\sigs^2-2\scrit}{2\sqrt{2}\sigs}\right]\right\}
\end{align}

\item \citet{Padoan11b}:

The critical density for  Equation \ref{eq:sfrff} for the Padoan \& Nordlund (2011, PN) model is their Equation 18:

\begin{align}
\label{pn.rhoc}
\rho_{crit,PN}/\rho_0=0.067\theta^{-2} \alphavir M_s^2 f(\beta)
\end{align}
where 
 $\theta \approx 0.35$ is the ratio
of the cloud size over the turbulent integral scale. The PN critical density therefore has the same proportionality with sonic Mach number and virial parameter as the KM model.
The effect of the magnetic field in slowing down star formation is encompassed in the function with dependency on plasma Beta ($\beta$, the ratio of thermal to magnetic pressure):
\begin{equation}
f(\beta) \equiv \frac{\left(1+0.925 \beta^{-3/2}\right)^{2/3}}{\left(1+\beta^{-1}\right)^2}
\end{equation}
and
the corresponding SFR$_{ff}$ with $\tff(\meanrho)/\tff(\rhocrit)$ pulled out of the integral in Equation \ref{eq:sfrff} is:

\begin{align}
\label{pnsfr}
SFR_{\rm{ff,PN}}=\frac{\eps}{2} \left\{1+\mathrm{erf}\left[\frac{\sigs^2-2\scrit}{2\sqrt{2}\sigs}\right]\right\}
\exp\left(\scrit/2\right)
\end{align}

The form of SFR$_{ff}$ in \citet{Padoan11b} therefore nearly identical the SFR in KM but multiplied by an additional
exponential term which arises from including $\tff(\meanrho)/\tff(\rho)=\tff(\meanrho)/\tff(\rhocrit)$. In both cases $\rho_{crit} \propto \alpha_{vir}M_s^2$.

\item \citet{Hennebelle11b} \& \citep{federrath12}:

The critical density for  Equation \ref{eq:sfrff} for the Hennebelle \& Chabrier (2011,HC) is specified in follow-up papers by \citet{federrath12} and \citet{Hennebelle13a}.  In the HC model, the critical density is defined by requiring that the turbulent Jeans length at the critical density is a fraction (which they denote as $ycut$) of the cloud size scale. This is in contrast to the KM and PN models in which  the critical density is proportional to ${M_s}^2$, implying that only very dense structures will lead to star formation. In the HC definition, any structure can collapse if its gravitational energy dominates over all sources of support (thermal, turbulent and magnetic), as long as the associated perturbation can grow and become unstable.
\begin{equation} \label{eq:scrit_hc}
\scrit{_\mathrm{,HC}} = \ln{\left[ \rho_\mathrm{crit,th}/\rho_0 + \rho_\mathrm{crit,turb}/\rho_0 \right]} \,,
\end{equation}
where the (magneto)thermal contribution is
\begin{equation} \label{eq:rhocrit_th}
\rho_\mathrm{crit,th}/\rho_0 \equiv \frac{\pi^2}{5}\ycut^{-2} \alphavir M_s^{-2} (1+\beta^{-1}) \,,
\end{equation}
and the turbulent contribution is
\begin{equation} \label{eq:rhocrit_turb}
\rho_\mathrm{crit,turb}/\rho_0 \equiv \frac{\pi^2}{15}\,\ycut^{-1}\,\alphavir \,.
\end{equation}
The HC model keeps the factor of $\tff(\meanrho)/\tff(\rho)$ inside the integral.  The SFR$_{\rm{ff,HC}}$ as reviewed in \citep{federrath12} is their Equation 40:

\begin{align}
\label{HCsfr}
SFR_{\rm{ff,HC}}=\frac{\eps}{2\phit} \left\{1+\mathrm{erf}\left[\frac{\sigs^2-\scrit}{\sqrt{2}\sigs}\right]\right\}\exp\left[(3/8)\sigs^2\right]
\end{align}

We note that the multi-free-fall SFR$_{\rm{ff}}$ extension discussed in \citet{federrath12} for the \citet{Krumholz2005} and \citet{Padoan11b} models has the same form as Equation \ref{HCsfr}. 
\end{itemize}

\begin{figure}
\includegraphics[width=9.8cm]{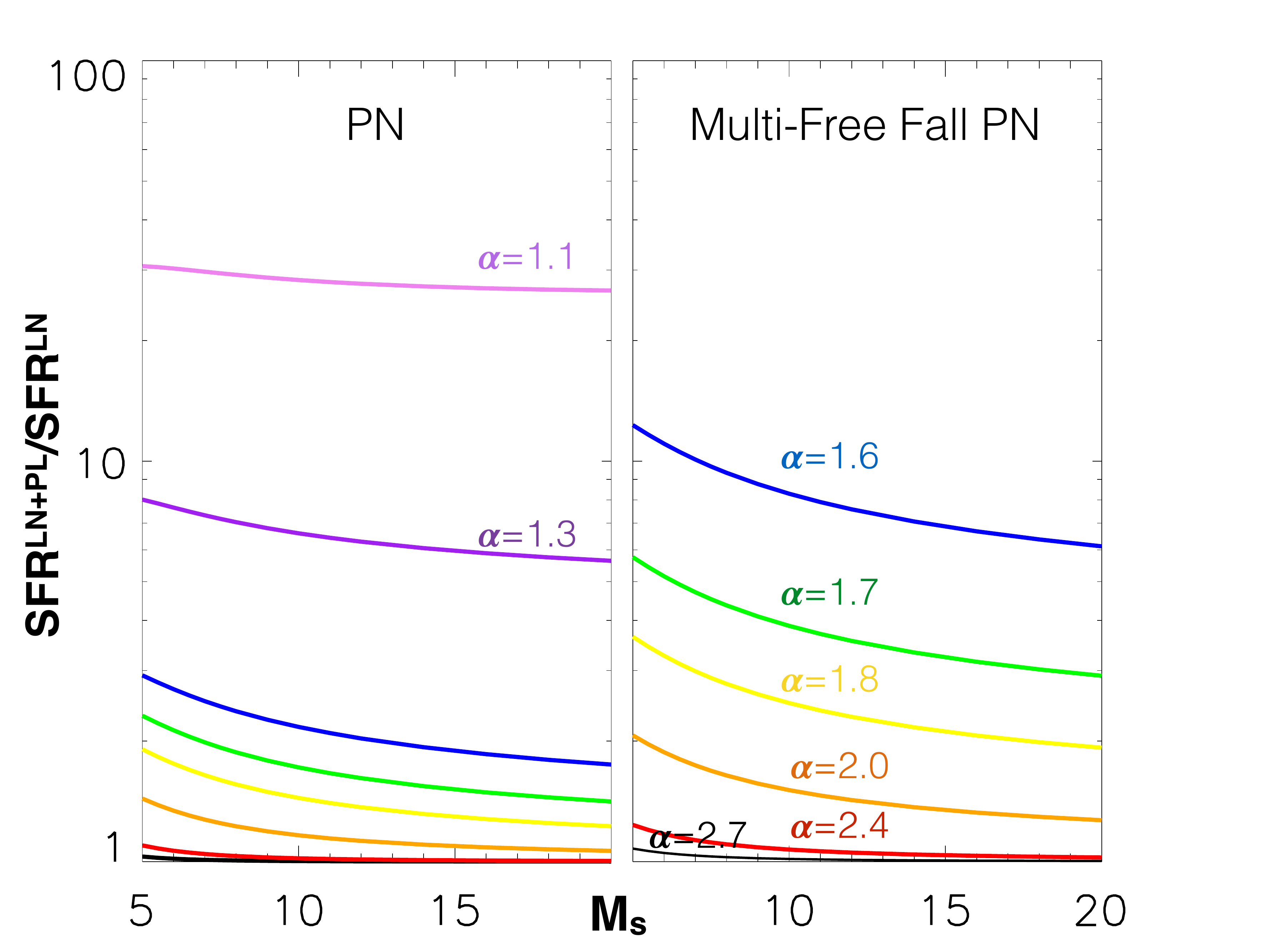}
\caption{
\label{fig:SFR_compare}
A comparison of the SFR$_{ff}$ for a piecewise lognormal and power law (LN+PL) density PDF and a lognormal (LN) density PDF (based on the critical density formula of \citet{Padoan11b}). We plot the ratio of $SFR_{ff}^{LN+PL}$/$SFR_{ff}^{LN}$  vs. the sonic Mach number for  b=0.4 (a mix of solenoidal and compressive forcing)  and allow the slope of the power law tail ($\alpha$) to vary (shown with different colored lines). 
Left panel: calculations done with the free-fall factor pulled out of the integral, i.e., Equations \ref{pnsfr} vs. \ref{sfrff_sol}. Right panel: calculations are multi-free-fall, Equations \ref{HCsfr} vs. \ref{sfrff_sol_mff}.
For $\alpha<2$ the LN+PL models show enhanced star formation rates over the pure lognormal calculations.
}
\end{figure}

\section{The Piecewise Lognormal + Power Law Density PDF}
\label{review2}

The analytic models for the SFR rate of \citet{Krumholz2005},~ \citet{Padoan11b}, and \citet{Hennebelle11b}, as well as follow-up modifications discussed by \citet{federrath12}, \citet{Federrath2015a}, and \citet{Gribel2017}, are all based on integrals over the lognormal density PDF thereby neglect contributions to the SFR that arise from the PDF power law tail. Numerous studies have now shown that the lognormal form of the PDF of column density or density describes the behavior of \textit{diffuse} molecular and atomic gas \citep{berkhuijsen08,Hill2008,burkhart10,Imara2016,Bialy2017}. However, as reviewed in the introduction, the dense star-forming gas PDF predominantly takes a piecewise lognormal plus power law form  in both 3D density (from simulations) and column density tracers \citep{Kainulainen09a,Collins12a,burkhartcollinslaz2015,burkhart15,schneider2015MNRAS.453L..41S,burkhart15,Imara2016}. In some cases the PDF of GMCs may be fully power law \citep{Lombardi2015AA,Alves2017AA}.

In light of these studies, we now consider the SFR calculation for a piecewise form of the density PDF in and around a star forming molecular cloud that consists of a lognormal at low density, a power law at high density and a transition point  ($s_t=ln(\rho_t/\rho_0)$) between the two.
This form of the PDF was considered in recent works such as \citet{Collins12a} and \citet{Burkhart2017ApJ...834L...1B}:

\begin{align}
p_{LN+PL}(s) = 
\begin{cases}
 N\frac{1}{\sqrt{2\pi}\sigs}e^{\frac{( s - s_0)^2}{2\sigs^2}}, & s < s_t \\
 N C e^{-\alpha s}, & s > s_t ,
\end{cases}
\label{eqn.piecewise}
\end{align}
where again, $\means=-\frac{1}{2}\,\sigs^2$

Here the normalization $N$ is determined by the
normalization requirement:
$N\int_{-\infty}^{\infty} p_{LN+PL}(s) ds=1$, and is given by
\begin{align}
\label{eq.norm}
N=\left( \frac{C e^{-\alpha s_t}}{\alpha} +\frac{1}{2}{+\frac{1}{2}\rm erf}\left(\frac{2s_t+\sigma_s^2}{2\sqrt{2}\sigs}\right) \right)^{-1}
\end{align}

We note that in our definition of the power law slope $\alpha$ is positive since the minus sign appears in the exponent separately.
Assuming that $p_{LN+PL}(s)$ is continuous and differentiable we can formulate an analytic estimate for $C$ and $\s_t$.  
These two conditions yield: 
\begin{align}
s_t&=(\alpha -1/2)\sigs^2
\label{eqn.st}
\end{align}

And the amplitude of the powerlaw:
\begin{align}
C=\frac{e^{1/2(\alpha-1)\alpha\sigma_s^2}}{\sigma_s \sqrt{2\pi}} 
\label{eqn.amplitude}
\end{align}

The transition density value between the lognormal and power law PDFs therefore depends on the  slope of the power law, the width of the lognormal and the mean density since $s_t=ln(\rho/\rho_0)$.

We pause here to discuss the physical meaning of the power law slope in the context of the piecewise PDF and the  transition density. 
 Numerical simulations of gravoturbulence suggest that the PDF of non-collapsing regions retrain the characteristics of 
the initial supersonic turbulence field (e.g. remain in the lognormal) while 
the density PDFs of collapsing
regions show a clear power law at high density \citep{Collins12a,Lee2015,Mocz2017}. Once the critical density for gravitational collapse is reached the power law begins to form.  The characteristic slope of that power law changes in roughly the cloud mean free fall time from steep, $\alpha \approx 3$, to shallow values, $\alpha\approx$1.5 to 1 \citep{Girichidis2014,Burkhart2017ApJ...834L...1B}. The exact time scale for $\alpha$ to saturate depends on the strength of the magnetic field \citep{burkhartcollinslaz2015}. The saturation value for the power law slope can be analytically determined from scale-free gravitational collapse \citep{Girichidis2014,Guszejnov2018}.
As $\alpha$ shallows, the transition density ($s_t$ between the PDF lognormal and power law moves towards lower density (Equation \ref{eqn.st}).  We illustrate the change of the power law slope and transitional density in Figure \ref{fig:PDF_theory}.

By combining Equation \ref{eqn.st} and Equation \ref{eqn.sigma}, the transition point can be further related to the physics of the medium \citep{Burkhart2017ApJ...834L...1B}: 
\begin{equation}
s_t=\frac{1}{2}(2\alpha-1)\ln[1+b^2M_s^2]
\label{eqn.rho0}
\end{equation}

\def\postshock{\rm{ps}}
The PDF transition density can  then be expressed in terms of
the isothermal post-shock density, $\rho_{\postshock}=\rho_0M_s^2$,
which is the density at which the turbulent energy density is equal to the thermal pressure:

\begin{equation}
1/2\rho_{\postshock}c_s^2=1/2\rho_0v^2 .
\end{equation}

Manipulating this relation we find that $M_s^2 = \rho_{\postshock}/\rho_0$, and Equation \ref{eqn.rho0} becomes
\begin{equation}
s_t=(\alpha -1/2)\ln[1+b^2\frac{\rho_{\postshock}}{\rho_0}].
\label{eqn.rho00}
\end{equation}

In the limit of strong collapse, $\alpha$ tends to 1.5 \citep{Girichidis2014} and therefore the $(\alpha -1/2)$ term is of order unity.

Therefore, 
\begin{equation}
\rho_t/\rho_0 \approx \left(1+b^2\frac{\rho_{\postshock}}{\rho_0}\right).
\label{eqn.rho2}
\end{equation}

For $\alpha=-1.5$  Equation \ref{eqn.rho2}  provides a direct physical interpretation for the PDF transition density in terms of the driving of the turbulence ($b$) and the post-shock density: $s_t(\alpha=1.5)=\sigma_s^2$. The transition density is proportional to the critical density for collapse based on the post-shock density derived in \citet{Krumholz2005} and \citet{Padoan11b}. 
We will explore the relationship between the transition density and the critical density for collapse in companion papers: Burkhart \& Mocz (2018) and Mocz \& Burkhart (2018).

\section{The Dynamic Star Formation Rate in the Gravoturbulent Media}
\label{SFR}

Given the analytic expressions for the normalization constants and transition density point (Equations \ref{eq.norm}-\ref{eqn.rho0}),  we can now integrate the piecewise density PDF and predict the star formation rate.

 In what follows, we consider the original formulation for the critical density for collapse $\rho_{crit}$ as defined in \citet{Padoan11b} in Equation \ref{pn.rhoc} with $\alphavir=1$, a weak magnetic field ($\beta=20$), and ratio of cloud size to turbulence integral scale of  $\theta=0.35$. 
In general, this is similar to the formulation of \citet{Krumholz2005}: $\rho_{crit,KM}/\rho_0=(\pi^2/15)\phit^2\times \alphavir M_s^2$, where $\alphavir=\phit=1$. We also consider the multi-free-fall piecewise density PDF SFR calculation derived in \citet{federrath12}, again using the critical density of \citet{Padoan11b} in Equation \ref{pn.rhoc}.
We plan to explore the full parameter space of $\alphavir$, $\beta$ and choice of critical density elsewhere.  

The integral to calculate the SFR$_{ff}$ now splits into an integral over the lognormal from  $\rho_{crit}$ to $\rho_t$ and an integral over the power law from $\rho_t$ to infinity:

\begin{align}
SFR_{ff}^{LN+PL} = \epsilon_0\mathlarger{\int}_{\scrit}^{\s_t}{\frac{\tff(\meanrho)}{\tff(\rho)} \frac{\rho}{\meanrho} \, p(s)_{LN} \deriv s}~\nonumber\\
+ \epsilon_0\mathlarger{\int}_{\s_t}^{\infty} {\frac{\tff(\meanrho)}{\tff(\rho)} \frac{\rho}{\meanrho} \, p(s)_{PL} \deriv s} \,.
\label{eq:sfrff_lnpl}
\end{align}

If we consider the free-fall factor weighting to be a constant $\tff(\meanrho)/\tff(\rho_{crit})$, and pull it out of the integral, then Equation \ref{eq:sfrff_lnpl}  evaluates to:
\begin{align}
SFR_{ff}^{LN+PL} &= {\rm{exp}(\scrit/2)}N\epsilon_0\bigg[\frac{1}{2}\mathrm{erf}\left(\frac{\sigs^2-2\scrit}{\sqrt{8\sigs^2}}\right)\nonumber\\
&-\frac{1}{2}\mathrm{erf}\left(\frac{\sigs^2-2s_t}{\sqrt{8\sigs^2}}\right)  + C\frac{\rm{exp}(s_t(1-\alpha))}{\alpha-1} \bigg]
\label{sfrff_sol}
\end{align}
which is limited to $\alpha > 1.0$
If the free fall factor is unity than the factor of ${\rm{exp}(\scrit/2)}$ is removed.

When the free-fall factor is kept inside the integral, the LN+PL star formation rate per free-fall time becomes:
\begin{align}
&SFR_{\rm{multiff}}^{LN+PL} = \frac{\rm{exp}(3\sigs^2/8)N\eps}{2} \bigg[\mathrm{erf}\left(\frac{\sigs^2-\scrit}{\sqrt{2\sigs^2}}\right)\nonumber\\
&-\mathrm{erf}\left(\frac{\sigs^2-s_t}{\sqrt{2\sigs^2}}\right) \bigg]
+ N\eps\left[C\frac{\rm{exp}(s_t(1.5-\alpha))}{\alpha-1.5} \right]
\label{sfrff_sol_mff}
\end{align}
which is limited to $\alpha > 1.5$

The SFR can be calculated according to Equation \ref{Sfrtrue}.
As shown in Figure \ref{fig:PDF_theory}, when $\alpha$ is steep, the density transition point is larger and hence the power law portion contributes very little to the $\sfrff$.
The power-law tail appears first at high densities
and steadily extends to lower densities as time proceeds, which has been shown analytically \citep{Girichidis2014} and numerically \citep{Collins12a,burkhartcollinslaz2015,Burkhart2017ApJ...834L...1B}.
Equation the $\sfrff$ depends increasingly on the power law portion of the piecewise PDF as $\alpha$ become shallower. 
%Additional parameters that affect the calculation of $\sfrff$, and which determine the transitional PDF density ($s_t$), are the width of the lognormal PDF $\sigs^2=\ln[1+b^2M_s^2]$, which depends on the turbulence driving ($b$) and sonic Mach number ($M_s$), and the mean density of the cloud ($\rho_0$).

In Figure \ref{fig:SFR_compare} we compare the $SFR_{ff}$ vs. sonic Mach number for a pure lognormal SFR$_{ff}$  (Equation \ref{pnsfr}) to the LN+PL $SFR_{ff}$ calculation. The left panel compares the original derivation from \citet{Padoan11b} to Equation \ref{sfrff_sol}, while the right panel compares the multi-free-fall expression derived in \citet{federrath12}, Equation \ref{HCsfr} to the LN+PL multi-free-fall expression in Equation \ref{sfrff_sol_mff}. 

We chose $b=0.4$ (a mix of solenoidal and compressive forcing) and allow the sonic Mach number (x-axis) and $\alpha$ to vary.  Different colored lines in Figure \ref{fig:SFR_compare} indicate different values of $\alpha$. For steep values of the power law tail slope, the $SFR_{ff}^{LN+PL}$ is very similar to the lognormal expression ($SFR_{ff}^{LN}$).  However, as $\alpha$ becomes increasingly shallow the star formation rate can grow to more than an order of magnitude above the pure lognormal calculation for both the multi-free-fall expression as well as the expression with the free-fall factor pulled out of the integrals.

\begin{figure*}
\includegraphics[width=18.8cm]{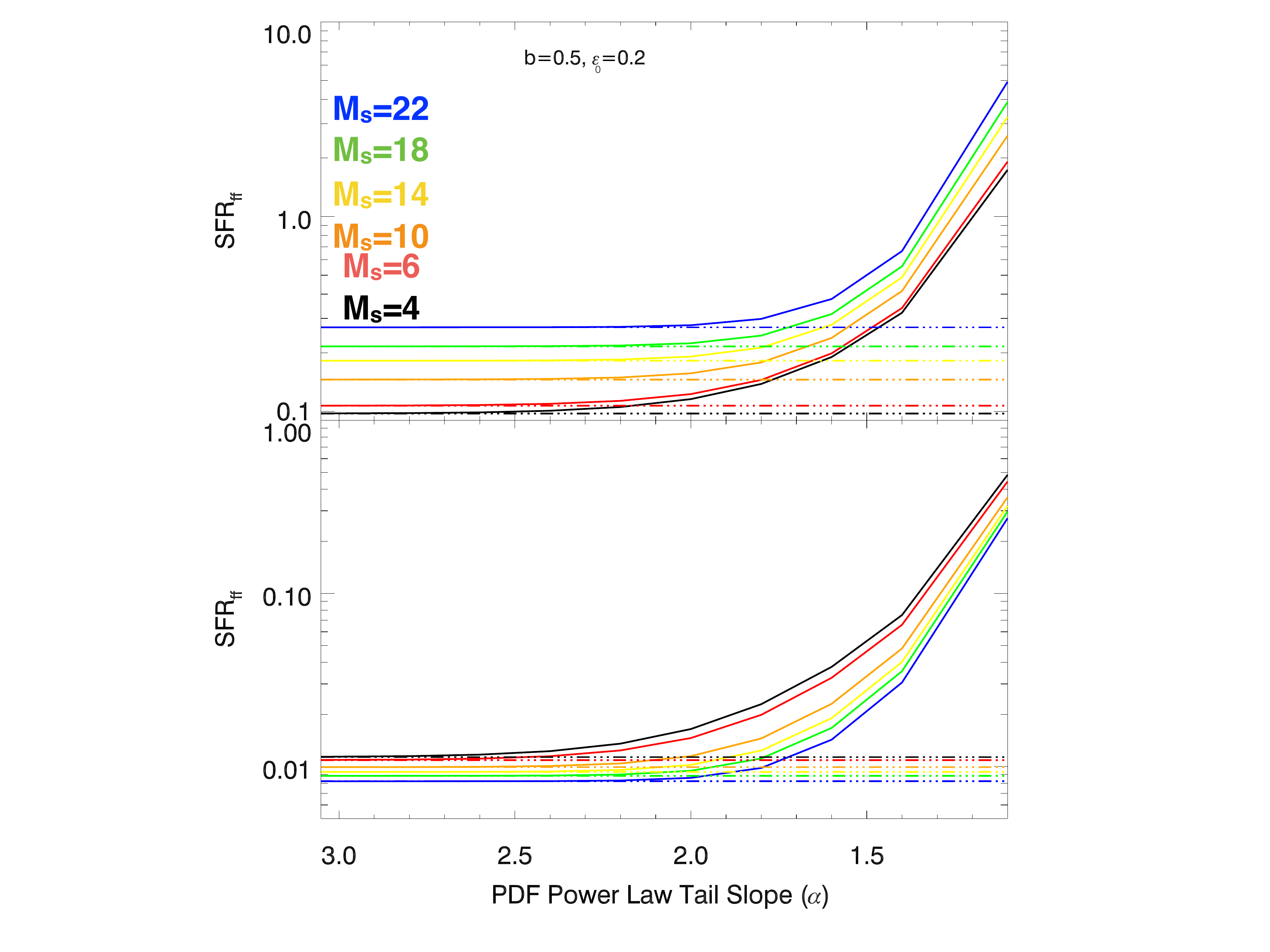}
\caption{
\label{fig:SFR_LNPL}
Analytic calculations for the star formation rate per free fall time (SFR$_{ff}$) vs. the density PDF power law slope ($\alpha$). The top panel using the free fall factor of \citep{Padoan11b}. Solid lines are for the SFR calculated using the lognormal+power law (LN+PL) model (Equation \ref{sfrff_sol}) and the straight dashed-dotted lines are for the pure lognormal (LN) models (Equation \ref{pnsfr}). 
The bottom panel shows the same but uses the free fall factor of \citep{Krumholz2005}, e.g. the factor of exp$(s_{crit}/2)$ is removed.
For both the top and bottom panel, the critical density used is the same (Equation \ref{pn.rhoc}).
Different colored lines indicate changing the sonic Mach number, with black representing the smallest value considered ($M_s=4$) and blue representing the largest value considered ($M_s=22$). 
The LN only SFR calculation only depends on the properties of the turbulence (sonic Mach number and forcing parameter $b$) and hence has no dependency on $\alpha$. Regardless of the free fall factor used, as $\alpha$ becomes increasingly shallow, the PL+LN model presented here begins to diverge from the LN model, with differences in the sonic Mach number becoming less significant. 
}
\end{figure*}

Figure \ref{fig:SFR_compare} also predicts that, when a time variable power law is included, the star formation rate is not constant. This is in contrast to the lognormal theories reviewed in Section \ref{review} that are constant with the properties of MHD turbulence, which set the width of the lognormal, e.g., Equation \ref{eqn.sigma}.

Figure \ref{fig:SFR_LNPL} shows the star formation rate per free fall time (SFR$_{ff}$) vs. the density PDF power law slope ($\alpha$) for both the LN+PL (Equation \ref{sfrff_sol}) and LN models. In the top panel, we compare the $SFR_{ff}^{LN}$ expression of \citet{Padoan11b}, Equation \ref{pnsfr}, with the $SFR_{ff}^{LN+PL}$ shown in  Equation \ref{sfrff_sol}.
The bottom panel shows the same comparison but using a free fall factor of unity as prescribed in \citet{Krumholz2005} (Equation \ref{sfrff_km}).
In both panels, the LN SFR calculation only depends on the properties of the turbulence (sonic Mach number and forcing parameter $b$) and has no dependency on $\alpha$. 
When using the \citet{Padoan11b} free fall factor (top panel), the SFR$_{ff}$ for both LN and LN+PL models increases with increasing sonic Mach number (i.e. increasing PDF width).  

The primary difference between the top and bottom panels stems from the factor of exp$(s_{crit}/2)$, which is present in the  \citet{Padoan11b} and absent in the \citet{Krumholz2005} model due to the different treatment of the free fall factor.
As noted in \citet{Krumholz2005} the combination of critical density increasing as sonic Mach number squared and a free fall factor of unity produces a SFR$_{ff}$ that is slightly anti-correlated with sonic Mach number, in contrast to the models of \citet{Padoan11b}, \citet{Hennebelle11b}, and \citet{federrath12}.  The Mach number dependency is also shown in Figure 1 of \citet{federrath12}. This demonstrates that different treatments of the free fall factor become important in both the LN and LN+PL models and can produce more than an order of magnitude difference in the value of SFR$_{ff}$. 

As $\alpha$ becomes increasingly shallow, the LN+PL models begin to diverge from the LN models. The width of the lognormal portion of the PDF sets the value of $\alpha$ for which the two models diverge. Lower values of $M_s$ show divergence at earlier stages of cloud evolution, i.e. steeper values of $\alpha$, regardless of the free fall factor criterion used. The models begin to diverge strongly around $\alpha=2$.

\section{Comparison With Observations}
\label{obs}

 In what follows, we compare the star formation rate using the LN+PL density PDF to observations of local GMCs.   We also investigate how our model can predict the slope and scatter of the relationship between the mass of molecular gas and the star formation rate \citep{Gao04a,Lada10a,Heiderman10a,Lada12a,Faesi2014}.
%We consider Equation \ref{sfrff_sol} and leave a full comparison of the effects of the multi-free-fall expression for future testing with numerical simulations.

%\begin{figure*}
%\includegraphics[width=14.8cm]{sfr_mass.pdf}
%\caption{
%\label{fig:SFR}
%The star formation rate (SFR) vs. the slope of the density PDF power law ($\alpha$) of the LN+PL model as compared to observations. Red, blue and black lines correspond to cloud masses of $10^4  M_{\odot}$, $10^3  M_{\odot}$ and $10^2 M_{\odot}$ respectively. The shaded regions in each color band mark the range of parameter space of sonic Mach number $M_s=5-25$ and $b=1/3-0.7$. All models assume a mean density of $\rho_0=500 cm^{-3}$ and $\epsilon=0.1$. 
%We overplot different local GMC SFRs and column density power law PDF slopes ($\alpha_{CD}$): Filled squares represent clouds from \citet{Lada10a} and \citet{Lombardi2015AA}. The observational points are color coded by reported cloud mass range \citep{Lada10a} above a threshold of A$_k=0.8$mag using NICEST.}
%\end{figure*}

\begin{figure*}
\includegraphics[width=18.8cm]{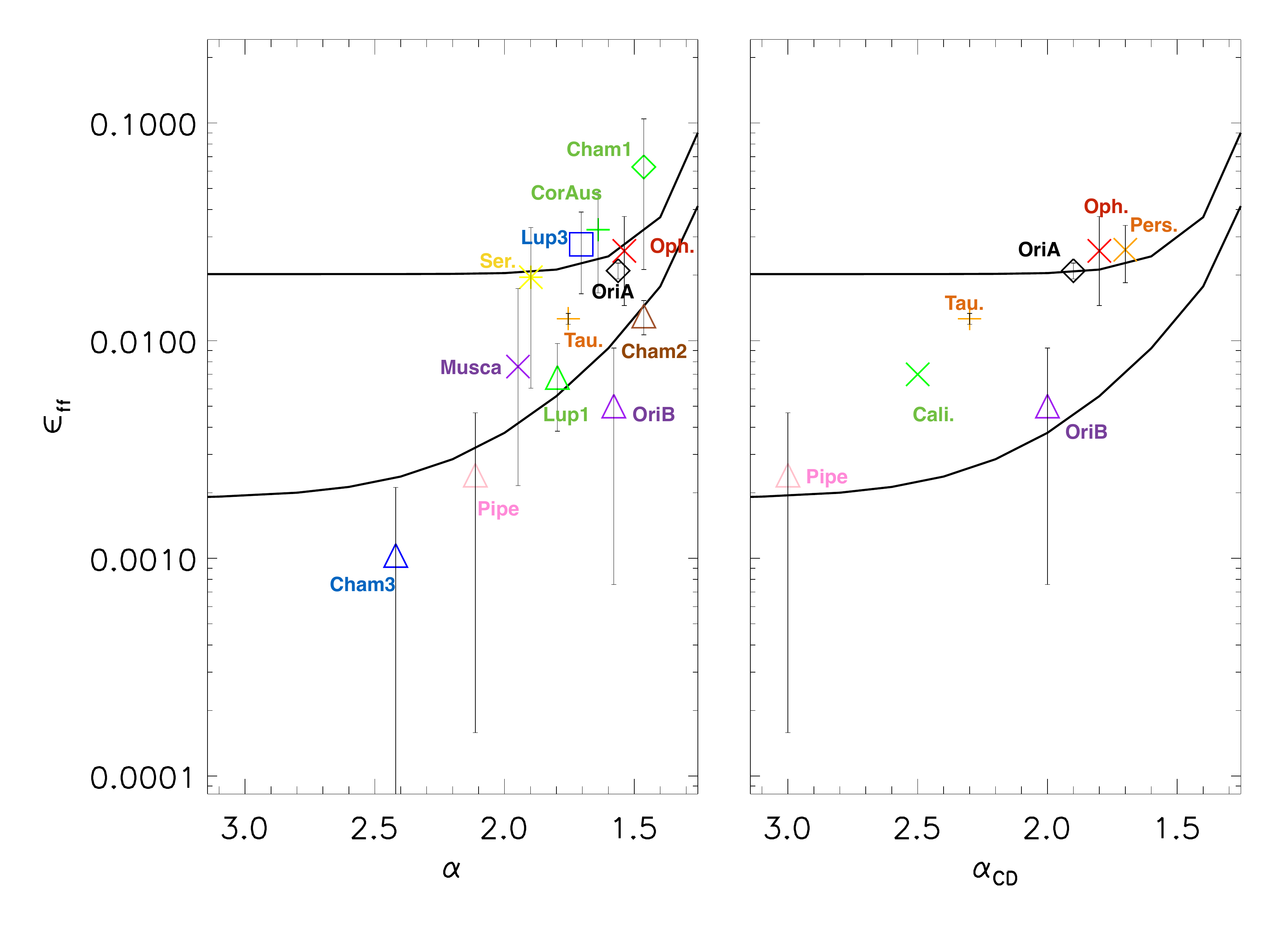}
\caption{
\label{fig:eps_ff}
The star formation efficiency per free fall ($\epsilon_{ff}$) vs. the slope of the density PDF power law (left panel) and column density PDF power law slope (right panel). The local GMC values for the PDF slopes, $<\epsilon_{ff}>$ and the references used are given in Table 1.  Local GMCs are represented by colored points and the LN+PL model predictions are shown in black lines bounding the parameter space of b=0.3-0.7 and M$_s=5-25$. }
\end{figure*}

\subsection{Local Giant Molecular Clouds and the Star Formation Efficiency}

\begin{figure*}
\includegraphics[width=15.8cm]{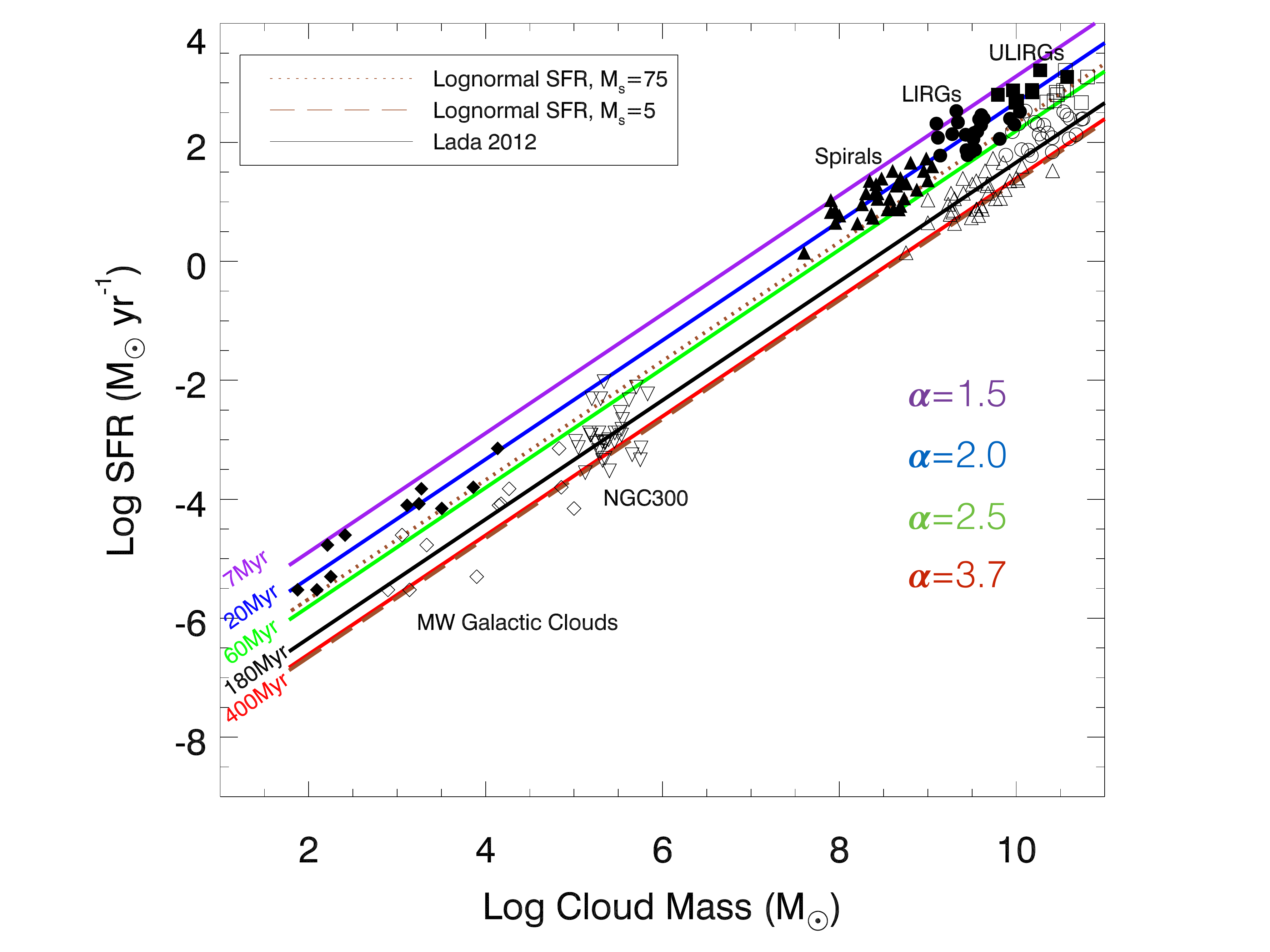}
\caption{
\label{fig:SFR_ks}
The SFR vs. cloud mass relationship, with the solid black line showing the relation from \citet{Lada12a}. \citet{Lada10a,Lada12a} published values of local GMCs (diamonds), spiral galaxies (upward triangles), Luminous Infrared Galaxies (LIRGS, circles), and Ultra-luminous Infrared Galaxies (ULIRG, squares). Individual GMCs from NGC 300 were published in \citet{Faesi2014} (downward triangles).   Open symbols correspond to diffuse gas tracers (either dust extinction at 0.1 mag or CO) while closed symbols indicate dense gas tracers (either dust extinction at 0.8 mag or HCN). Rainbow colored solid lines are different LN+PL model predictions for the SFR with $M_s=5, b=1/3$ and different values of $\alpha$, which can greatly change the amplitude of the SFR and hence contribute to the scatter in SFR despite the constant turbulence properties. For each of the LN+PL models we over-plot the corresponding depletion times. Two lognormal only models are shown in brown dotted ($M_s=75, b=1/3$)  and dashed lines ($M_s=5, b=1/3$) can only explain the scatter if they have extreme variations in turbulence parameters or the ad-hoc SFE parameter. }
\end{figure*}

A number of observational studies have been conducted to understand the relationship between the dense gas mass fraction, the gas density PDF and the properties of star formation such as the star formation efficiency per free fall time ($\epsilon_{ff}$).
We define $\epsilon_{ff}$ as in \citet{Krumholz2005}:
\begin{equation}
\epsilon_{ff}=SFR\frac{t_{ff}(\rho_0)}{M_{\rm{cloud}}}
\end{equation}
Examples of such studies include \citet{Lada10a}, \citet{Heiderman10a} and \citet{Kainulainen2014}, who investigated the level of star formation activity and cloud structure in nearby GMCs. \citet{Krumholz12a} summarized the observed data from 
\citet{Lada10a} and \citet{Heiderman10a} and derived the star formation efficiency per free fall time (see their Table 2). 
%They used a uniform set of infrared extinction maps to measure the cloud mass and compared these with inventories of young stellar objects (YSO) within the clouds.  They showed that the star formation rate based on the number of YSOs (NYSO) is: SFR=$0.25\times NYSO\times 10^{-6} \rm{M_{\odot}yr^{-1}}$. The SFR based on NYSO can vary considerably in local clouds, independent of their mass and size.  
For many of the clouds studied in \citet{Lada10a}, \citet{Lombardi2015AA} also calculated the slope of their column density PDF power law ($\alpha_{CD}$) using dust extinction/emission data from \textit{Planck}, \textit{Herschel} and 2MASS. 
\citet{burkhartcollinslaz2015} also noted that there is a correlation in local GMCs with shallower column density power law tails and an increase in the number of young stellar objects (NYSO), which is proportional to the SFR.  More recently,  \citet{Kainulainen2014} derived 3D density GMC PDFs for 16 clouds using a wavelet-based reconstruction technique.  The reported cloud masses, cloud densities, and NYSO, all of which can be used to derive the SFRs and SFEs.  \citet{Kainulainen2014}  also measured the exponent of the radial density distribution ($\kappa$), which is related to the PDF density slope as $\kappa=3/\alpha$.  
We compile values of $\alpha$, $\alpha_{CD}$ and the average value of $\epsilon_{ff}$ from the above mentioned studies and present them in Table 1.

\begin{table}
\caption{GMC star formation efficiency per free fall time, references: $^{1}$\citet{Kainulainen2014},$^{2}$\citet{Lada10a}, $^{3}$\citet{Heiderman10a},$^{4}$\citet{Krumholz12a},$^{5}$\citet{Lombardi2015AA}   }
\begin{center}
\begin{tabular}{ccccc}
\hline
\hline
GMC & $<\epsilon_{ff}>$ & $\alpha_{CD}$ & $\alpha$ & Ref.\\
\hline
Orion A &   0.021   & 1.9  &  1.56& 1,2,4,5\\
Orion B &  0.005  & 2 &  1.58& 1,2,4,5  \\
Oph.    & 0.026   & 1.8  & 1.54& 1,2,3,4,5  \\
Taurus &0.013    &   2.3 &   1.75&  1,2,4,5 \\
Serpens & 0.017  & N/A &   1.9& 1,3,4 \\
ChamI & 0.06 & N/A  &   1.46& 1,3,4 \\
ChamII & 0.013    &N/A  &   1.46& 1,3,4 \\
LupI & 0.007 &  N/A &   1.8& 1,2,3,4 \\
LupIII & 0.028  & N/A &   1.7& 1,2,3,4 \\
CorA &0.032 & N/A  &   1.64& 1,2,3,4 \\
Pipe & 0.002 &  3 &   2.11&  1,2,4,5\\
Musca & 0.008 &N/A   &   1.95&  1,3,4\\
ChaIII &0.001 &   N/A &    2.42& 1,3,4 \\
Perseus & 0.026&  1.7 & N/A& 2,3,4,5 \\
Cali. & 0.0065&  2.5 & N/A& 2,4,5 \\
\hline
\hline
\end{tabular}
\end{center}
\label{tab:data}
\end{table}

We plot $\epsilon_{ff}$  vs. power law slope in Figure \ref{fig:eps_ff}.
Error bars on the observations are computed from the standard deviation of the different reported values of $\epsilon_{ff}$ in the literature.  The left panel shows $\epsilon_{ff}$ vs. the 3D density $\alpha$ reported in \citet{Kainulainen2014} for local GMCs.  The right panel presents $\epsilon_{ff}$ vs. the column density PDF slopes published in \citet{Lombardi2015AA}.
For both density and column density PDF power law slopes we note a clear trend of rising $\epsilon_{ff}$ with shallower power law tail slope in the observations. This trend is within the expectations of the LN+PL model (black lines) range for sonic Mach numbers of 5-25 and $b$ values of $1/3-0.7$, although we emphasize that the amplitude depends primarily on the feedback parameter ($\epsilon_0$) and which version of the free fall factor is employed.  Regardless of these, the rise with shallower $\alpha$ is expected from the model and observed in the local GMCs.

We confirm that there are line of sight effects that may produce a difference between the actual density PDF $\alpha$ and the measured column density PDF $\alpha_{CD}$.  These LOS effects may also apply to the reconstructed density power law slope measured in \citet{Kainulainen2014}. \citet{burkhartcollinslaz2015} and \citet{Burkhart2017ApJ...834L...1B} studied the column density PDF properties and compared them to the density PDF properties from \citet{Collins12a}. They found that for the same time step and plasma $\beta$ (mean magnetic field strength) the power law slopes of column density  PDF can change slightly depending on the LOS orientation.
\citet{Girichidis2014} and  \citet{Guszejnov2018} performed analytic calculations for the evolution of the power-law PDF slopes for both density and column density of all
material which undergoes complete fragmentation (e.g. goes
on to form stars). In particular, the results of \citet{Guszejnov2018} suggest that the density and column density power law slopes should converge to values of -1. 
However, they argue that the column density PDF is
affected by LOS effects which steepens the slope, most likely producing final slopes roughly between -1 and -2.  Indeed for all the clouds examined here $\alpha_{CD}>\alpha$.
Additional studies comparing the analytic results of \citet{Girichidis2014} and \citet{Guszejnov2018} to numerical simulations will be useful to determine the correspondence between column density PDF $\alpha_{CD}$ and the density PDF $\alpha$.

Despite the observational complications of measuring $\epsilon_{ff}$ and the PDF properties \citep{Alves2017AA}, a very clear trend of increasing $\epsilon_{ff}$ with shallower $\alpha$ is seen for the local GMCs.
An important distinction between the analytic SFR model derived in this work and the works of \citet{Krumholz2005}, \citet{Padoan11b} and \citet{Hennebelle11b} is that they provide a constant $SFR_{ff}$ and/or $\epsilon_{ff}$ for given turbulence parameters, such as the sonic Mach number and the forcing, which set the lognormal width.  However, it is not likely that differences in $\epsilon_{ff}$ observed in the data is due the properties of turbulence, as most of these clouds have similar sonic Mach numbers and similar line width size relations \citep{Kainulainen2017}.
The LN+PL model $SFR_{ff}$ calculation presented here is inherently time varying as the slope of the power law varies with cloud evolution due to gravitational collapse and feedback.  The level of turbulence is secondary to the $SFR_{ff}$ when the power law is taken into account (see Figure 3).

%As $\alpha$ becomes increasingly shallow with time, the $SFR_{ff}$ increases, as has been observed in both observations \citep{Palla00a,schneider15,burkhartcollinslaz2015,Lee2016,Caldwell2018MNRAS.474.4818C} and in simulation \citep{Collins12a,federrath12,burkhart15,Lee2015ApJ...800...49L,Lee2017arXiv171100316L,Mocz2017}. $\alpha$ should also re-steepen as star formation feedback becomes important.

\subsection{From Galactic to Extragalactic: Dense Gas Mass and the Star Formation Rate}

%Local Milky Way GMCs provide an excellent laboratory for testing the predicted relationship between the SFR and the density PDF. However, investigations of nearby galaxies can provide an additional large scale view of the association between molecular gas and star formation. 

A number of insightful observational surveys have shed light on the relationship between ISM properties and extragalactic SFR scaling laws.
\citet{Gao04a} found a linear relation between the SFR traced by total infrared luminosity and the luminosity in a dense gas tracer (HCN)
averaged over entire galaxies (see also, \citet{Wu05a}).
 Similar relations between dense gas mass or gas surface density and star formation rates have been observed in a number of recent extragalactic surveys \citep{Bigiel2008,Schinnerer13a,Pety13a,Hughes13a,Colombo14a,Faesi2014,Usero2015,Bigiel2016,Leroy2017,Federrath2017} as well as in local well-resolved Milky Way GMCs \citep{Heiderman10a,Lada10a}  
 
 %found a linear relationship between 
%the SFR and total dense molecular gas (defined in that study as gas with $n_{H2} > 10^4\rm{cm}^{-3}$) in a sample of ten well-resolved nearby molecular clouds.

The relationship between molecular gas mass and the SFR was further extended in \citet{Lada12a} and \citet{Faesi2014}, who showed that the same linear correlation smoothly extends across more than five orders of magnitude when connected to local galaxy dense gas masses and star formation rates and is consistent with the Gao \& Solomon (2004) and Wu et al. (2005) results to within a factor of three. 
One interpretation of this linear slope is an approximately
constant depletion time  ($\rm{t_{\rm{depl}}(yr)=M_{\rm{mol}}(M_{\odot})/SFR (M_{\odot}/yr)}$), i.e.,  the timescale for the molecular gas  to be converted into stars. Extragalactic studies consistently derive average depletion times of about 2 Gyr \citep{Bigiel2008,Leroy13a,Leroy13b}, however deviations have been noted \citep{Leroy11a,Saintonge11a,Saintonge11b,Sandstrom13a,Burkert2013ApJ...773...48B}.  These deviations could be due to systematics and/or true environmental differences in galaxies. Furthermore, local MW GMCs and resoled GMCs in nearby galaxies have a median $t_{\rm{depl}}$ of 180-230 Myrs, an order of magnitude shorter than the typical values found in unresolved extragalactic studies \citep{Lada10a,Lada12a,Faesi2014}. Dense gas tracers have even shorter depletion times of around 20 Myrs \citep{Lada12a}.

We explore these trends using the predictions of LN+PL SFR model and compare with the LN only model.
Figure \ref{fig:SFR_ks} shows the SFR vs. cloud mass for local GMCs and galaxies, with the solid black line showing the fit from \citet{Lada12a} and data from \citep{Gao04a}. \citet{Lada10a,Lada12a} published values of local GMCs (diamonds), spiral galaxies (upward triangles), Luminous Infrared Galaxies (LIRGS, circles), and Ultra-luminous Infrared Galaxies (ULIRG, squares).  GMCs from NGC 300 were published in \citet{Faesi2014} (downward triangles). Opened symbols correspond to diffuse gas tracers (either dust extinction at 0.1 mag or CO) while closed symbols indicate dense gas tracer (either dust extinction at 0.8 mag or HCN). Rainbow colored solid lines are different LN+PL model predictions for the SFR with $M_s=5, b=1/3$ and different values of $\alpha$, which can greatly change the amplitude of the SFR and hence contribute to the scatter in SFR despite the constant turbulence properties. For each of the LN+PL models we over-plot the corresponding depletion times. Two lognormal only models are shown in brown dotted ($M_s=75, b=1/3$) and dashed lines ($M_s=5, b=1/3$).

The lognormal and steep LN+PL models (red, orange and yellow lines) show agreement with observations of diffuse molecular gas tracers (open symbols).  These tracers are more likely to lie within the lognormal portion of the gas PDF since they are diffuse.  However the high density gas tracers (e.g. HCN, filled symbols) show better correspondence with shallower power law slope values of the SFR (green, blue and purple lines). The LN SFR models could explain these points as well but only for extreme sonic Mach numbers in the range of $M_s=75-100$ \citep{Federrath13c,Usero2015,Federrath2017} or invoking large variations in the star formation efficiency. Typical sonic Mach numbers measured in local star forming Milky Way clouds lie in the range of $M_{\rm{s,Milky Way}}=4-20$, \citep{burkhart10,Kainulainen13a,Kainulainen2017}. 

As seen in the previous section in Figure \ref{fig:SFR_compare}, the LN+PL model SFR can change by more than an order of magnitude for different values of $\alpha$ and hence can explain the scatter in the SFR with the average observed Mach numbers in the ISM. A shallower value of  $\alpha$ corresponds to increasing the fraction of dense gas (f$_{dense}$), which is representative of star formation law discussed in \citet{Lada12a}: SFR $\propto f_{dense}M_{cloud}$.

For each of the three LN+PL models we over-plot the corresponding depletion times, $t_{\rm{depl}}=7, 20, 60, 400$ Myr for $\alpha=1.5,2.0, 2.5, 3.7$, respectively. Clouds with shallow power law tails have the shortest depletion times, the most dense self-gravitating gas in the power law tail, and rapid star formation, while clouds with steep power law tails or lognormal distributions with moderate sonic Mach numbers have longer depletion times.

The star formation rate and depletion time is primarily controlled by the local dense gas fraction which, in the context of the density PDF SFR model presented here, is characterized by the amount of material in the power law portion of the PDF. Once the critical density is reached, the value of  $\alpha$ rapidly shallows and the lognormal portion of the SFR calculation becomes insignificant. Therefore the value of $f_{dense}$, and hence the overall SFR, is determined by the competition of gravity vs. feedback and secondarily on magnetic fields and large scale turbulence.

\section{Discussion}
\label{discussion}

The LN+PL model discussed here can be retermed as \textit{a gravoturbulent star formation model} since the development of lognormal plus power law density distribution is 
consistent with the picture of a turbulent molecular cloud undergoing 
gravitational collapse. 
Both numerical experiments and observations point to a picture of \textit{a dynamic star formation rate}, which increases in time as clouds become gravitationally unstable \citep{Hartmann01a,Vazquez-Semadeni07a,Heitsch08a,Heitsch08b,hennebelle08a,Banerjee2009,Zamora-Aviles12a,Burkert2013ApJ...773...48B,Girichidis2014,Zamora-Aviles14a} and later halts as feedback/supernova destroy the star forming region.
The model presented here, which includes both the lognormal and power law components, is inherently time varying as the power law slope is expected to change as the cloud contracts and later is destroyed by feedback.  Previous studies have investigated the SFR or IMF from a power law only distribution with single values of $\alpha$ \citep{Collins12a,Girichidis2014,Lee2017arXiv171100316L}. However, GMCs have different observed values of the power law slope and  $\alpha$ is expected to become shallow as the cloud evolves from rapidly collapsing filaments \citep{Collins12a,federrath12,Federrath13c,schneider15,burkhart15,Lee2015ApJ...800...49L,Lee2017arXiv171100316L,Mocz2017,Caldwell2018MNRAS.474.4818C} and the $SFR_{ff}$ increases.
  $\alpha$ may steepen again as feedback from OB stars destroy the GMC as maybe indicated in GMCs which show multiple power law tail slopes \citep{schneider2015MNRAS.453L..41S}. 
  
As mentioned above, an important distinction between the analytic model presented here and the SFR models of \citet{Krumholz2005}, \citet{Padoan11b} and \citet{Hennebelle11b} is that they provide a constant $SFR_{ff}$ for a given sonic Mach number (i.e. lognormal width), $\beta$, $\alpha_{vir}$ and $\epsilon_0$. The model presented here provides the same value of the SFR when the power law tail is very steep and the PDF is dominated by the lognormal.  However for clouds with prominent high density power law PDFs the SFR is observed to increase and our model accounts for this effect \textit{without the need to invoke stronger turbulence or alternative environmental parameters}.

%A follow up study (Paper 2) will explore our model with the full parameter space of  $\rho_{crit}$, $\beta$, $\alpha_{vir}$, $\epsilon$ and include testing with numerical simulations to determine how the multi-free-fall expression (Equation \ref{sfrff_sol_mff}) compares with the constant free-fall factor (Equation \ref{sfrff_sol}). 

The model presented here suggests that star formation rates and efficiencies are very sensitive to the increasing influence of the gas self-gravity and feedback. The key observable is the manifestation of the power law in the density PDF.  In that sense we can explain why observed star formation rates do not show sensitivity or strong correlation with local measurements of velocity dispersion or second moment maps \citep{Leroy2017} and why star formation correlates strongly with dense gas fraction.  We will explore this point further in a follow up paper (Burkhart \& Mocz 2018, submitted). 

%The model here suggests that the simple cloud structure produced by turbulence, gravity and feedback controls the 
%protostellar population and thus the global SFR in the cloud. 

We emphasize that in the analytic formulation for the SFR and SFE outlined in this work there is \textit{no need to invoke the ionization state of the cloud or complex plasma processes such as the Hall effect, or ambipolar diffusion.} Turbulence, turbulent magnetic fields, and gravity set the basic initial conditions for the star formation rates and efficiencies. 
Turbulence sets the initial density distribution (i.e. with a lognormal PDF) and can also control the fluctuation, amplification, and diffusion of the magnetic field. A process known as reconnection diffusion \citep{lazarianvishniac99,LEC12} is mediated by the properties of turbulence. Numerical simulations have shown that magnetic reconnection diffusion can be an important process for mediating collapse \citep{Santos-Lima12a,LEC12,Mocz2017} especially in clouds with sub-Alfv\'enic and magnetically sub-critical initial conditions \citep{Crutcher2009,Crutcher10a,Crutcher12a}.  In this paper we only explored a weakly magnetized regime of parameter space with our choice of $\rho_{crit}$ but plan to explore the effects of strong magnetization, where reconnection might play a more important role, in a future work.

The gravoturbulent star formation model has implications for both the Kennicutt-Schmidt relationship as well as the numbers of YSOs in local GMCs. The analytic model outlined here suggests that star formation scaling laws from galaxy scales to cloud scales are driven primarily by variation in the amount of dense gas (set by the power law slope $\alpha$) and secondarily on the properties of turbulence in the clouds.
This is in agreement with recent observational work by \citet{Lada2017} on the California GMC as well as work by \citet{Elmegreen2018} on explaining scatter in the Kennicutt-Schmidt relation.
The gravoturbulent model can explain the scatter in the SFR vs. dense gas mass observations \citep{Heiderman10a,Lada10a,Lada12a,Faesi2014,Leroy2017}, without the need to invoke extremely large turbulent energies (e.g. large $M_s$). 
The gravoturbulent model SFR changes by more than an order of magnitude for different values of $\alpha$ and hence could be an explanation for the scatter in the SFR in both local GMCS as well as in extragalactic quiescent vs. starbursting systems.  Clouds with shallow power law tails have SFRs dominated by self-gravity and we predict they also have higher dense gas fractions, shorter depletion times and rapid star formation.  This is in agreement with the observational studies of \citet{Usero15a} and \citet{Leroy2017}, who found that gas with stronger self-gravity forms stars at a higher rate (lower depletion time). Moreover, these studies found that the star formation efficiency is anti-correlated with the velocity dispersion (Mach number), which is in tension with some lognormal turbulence theories of star formation.

The star formation theory derived in this paper is suitable for application to galaxy/cosmological simulation sub-grid models \citep{Hopkins2013MNRAS.432.2647H}. Our model predicts that star formation depends primarily on the amount of dense self-gravitating gas and secondarily on the properties of ISM turbulence, which set the initial density fluctuations.  Gravitational collapse primarily controls star formation until it is shut of by mass loss and/or feedback processes. 
Galaxy formation simulations which seek to include ISM physics should adopt a SFR law which has a primary dependency on a dense gas probability and secondarily on the conditions of the ISM including, $\alpha$, $M_s$, $\alpha_{vir}$ and $\beta$.

\section{Conclusions}
\label{conclusions}

Stars form in supersonic turbulent molecular clouds that are self-gravitating. 
We presented an analytic determination of the star formation rate in a gravoturbulent medium based on the density probability distribution function of molecular clouds having a piecewise lognormal \textit{and} power law (LN+PL) form. Previous analytic models for the star formation rate based on the density distribution used only the lognormal portion of the PDFs, with properties determined by supersonic turbulence, and ignored the contribution of the high density power law tail set by self-gravity.

In the gravoturblent model presented here, the star formation rate per free-fall time is calculated via an integral over the lognormal from a critical density ($\rho_{crit}$) to the transition density between the lognormal and power law ($\rho_t$) and then an integral over the power law from $\rho_t$ to infinity. This accounts for clouds which are in the early stages of collapse to later evolutionary stages.  The transitional density is analytically determined from the conditions of continuity and differentiability of the PDF \citep{Collins12a,Burkhart2017ApJ...834L...1B}.  We find that:

\begin{itemize}
\item  As the slope of the power law tail becomes shallower the SFR$_{ff}$ in the gravoturbulent model increases past the expectations of the turbulence SFR models (i.e. pure lognormal PDF models).

%\item The gravoturbulent model of star formation is based on a density PDF where turbulence coupled with gravitational collapse sets the initial conditions for the formation of stars.  The properties of turbulence set the initial density distribution however the gas self-gravity produces the largest contribution to SFR$_{ff}$. This is because gravity sets the excess from a lognormal distribution in the high density power law portion of the PDF. 

\item We show that star formation efficiency per free fall time in local giant molecular clouds increases with shallower PDF power law slopes, in agreement with our model. 

\item The approach presented here  can explain why star formation is spatially and temporally variable within a cloud and can accelerate in pace despite roughly constant ISM turbulent energy and magnetic field strengths.

\item Clouds with shallow power law tails have a higher dense gas fraction and therefore short depletion times and rapid star formation.

\item Our model can explain both star-bursting and quiescent star-forming systems without the need to invoke extreme variations in the local interstellar environment or large sonic Mach numbers. 

\end{itemize}

\acknowledgments
B.B. is grateful for valuable discussions with Shmuel Bialy, Thomas Bisbas, Hope Chen, David Collins, Christopher Faesi, Christoph Federrath, Adam Ginsburg, Alyssa Goodman, Jouni Kainulainen, Mark Krumholz, Charlie Lada, Alex Lazarian, Adam Leroy, Abraham Loeb, Christopher Mckee, Anna Rosen, Zachary Slepian, Amiel Sternberg, and Catherine Zucker. B.B. acknowledges support from the Institute for Theory and Computation (ITC) Harvard-Smithsonian Center for Astrophysics Postdoctoral Fellowship.

\end{document}